# The role of alcohol outlet visits derived from mobile phone location data in enhancing domestic violence prediction at the neighborhood level

Ting Chang [a], Yingjie Hu [a], Dane Taylor [b], and Brian M. Quigley [c]

[a] Department of Geography, University at Buffalo, The State University of New York, USA
[b] Department of Mathematics, University at Buffalo, The State University of New York, USA
[c] Department of Medicine, University at Buffalo, The State University of New York, USA

**Abstract:** Domestic violence (DV) is a serious public health issue, with 1 in 3 women and 1 in 4 men experiencing some form of partner-related violence every year. Existing research has shown a strong association between alcohol use and DV at the individual level. Accordingly, alcohol use could also be a predictor for DV at the neighborhood level, helping identify the neighborhoods where DV is more likely to happen. However, it is difficult and costly to collect data that can represent neighborhood-level alcohol use especially for a large geographic area. In this study, we propose to derive information about the alcohol outlet visits of the residents of different neighborhoods from anonymized mobile phone location data, and investigate whether the derived visits can help better predict DV at the neighborhood level. We use mobile phone data from the company SafeGraph, which is freely available to researchers and which contains information about how people visit various points-of-interest including alcohol outlets. In such data, a visit to an alcohol outlet is identified based on the GPS point location of the mobile phone and the building footprint (a polygon) of the alcohol outlet. We present our method for deriving neighborhood-level alcohol outlet visits, and experiment with four different statistical and machine learning models to investigate the role of the derived visits in enhancing DV prediction based on an empirical dataset about DV in Chicago. Our results reveal the effectiveness of the derived alcohol outlets visits in helping identify neighborhoods that are more likely to suffer from DV, and can inform policies related to DV intervention and alcohol outlet licensing.

---

## 1. Introduction

Domestic violence (DV) is a serious public health issue, with 1 in 3 women and 1 in 4 men experiencing some form of partner-related violence every year (Black et al., 2011). In the United States, the economic cost of domestic and family violence is estimated to be over $12 billion per year, and about 10 million people are affected by DV every year (Huecker et al., 2021). In the UK, the cost of domestic violence to society is estimated to be around £15.7 billion per year (Walby,

2009). DV can lead to various consequences affecting the victims for their entire lifespan, including but not limited to physical injuries, mental health issues, and loss of productivity (Beyer et al., 2015; Felker-Kantor et al., 2017). Given its substantial costs to individuals, society, and the economy, reducing and preventing DV are critical tasks for policy makers and related organizations worldwide (Anderson, 2021).

Over thirty years of research has shown that excessive alcohol use is one of the strongest predictors of DV (Leonard & Quigley, 2017). The link between alcohol and DV has been established across countries (Abramsky et al., 2011), in general population surveys (Quigley et al., 2018), clinical samples (O'Farrell et al., 2004), and confirmed in meta-analyses (Foran & O'Leary, 2008). Despite their strong association, existing studies examining alcohol use and DV were typically conducted at the individual level (or couple level). Much fewer studies were done at the neighborhood level (Beyer et al., 2015). While individual-level studies improve our understanding of the social, behavioral, and psychological factors associated with the individuals involved in DV, neighborhood-level studies can help us identify the neighborhoods where DV is more likely to happen and inform neighborhood-level policies. For example, DV intervention and prevention programs could consider distributing more educational materials about alcohol use and DV to those neighborhoods that are more likely to suffer from alcohol-related DV.

One challenge that hinders the study of alcohol use and DV at the neighborhood level is the lack of neighborhood-level alcohol use data. Consequently, we do not know, e.g., the general level of alcohol use of the residents living in different neighborhoods. While it is possible to use surveys or interviews to collect such data, they require a large amount of financial and labor resources. This is especially the case when we need to collect data for many neighborhoods in a large geographic area (e.g., a major metropolitan area). Even when the required resources are available, completing such a survey can take a long time. The collected data could also be affected by the training of the interviewers (Schafer et al., 2004; Abramsky et al., 2011), since people may not self-report accurate alcohol use information due to its sensitivity.

The emergence of anonymized mobile phone location data offers new possibilities for measuring variables indicative of alcohol use at the neighborhood level. This type of data typically provides information about how people move around in a geographic area (e.g., a city) and how people interact with different places, called points-of-interest (POIs) (McKenzie et al., 2015; Gao et al., 2017; Liang et al., 2020; Sparks et al., 2020; Gao et al., 2020). In addition to gas stations, parks, coffee shops, and other places, these POIs also include various alcohol outlets, such as liquor stores, bars, breweries, and wineries. Thus, it is possible to derive information about how people visit various alcohol outlets and the home locations of the anonymous visitors from such mobile phone location data. To further protect individual privacy, the data providers generally do not provide the exact home locations of the individuals in their data but aggregate these locations to a certain level of geographic units, such as neighborhoods. Utilizing such anonymized mobile phone location data, we can derive information about how the residents of a neighborhood visit various alcohol outlets during a time period.

It is worth noting that mobile phone location data do not contain actual purchase information, i.e., we do not know whether a person made a purchase at a POI and what items were purchased. Accordingly, the derived alcohol outlet visits are not equivalent to alcohol purchase or alcohol use. However, they are related to each other. In a recent study, we examined the relation between alcohol sales and people's visits to four types of alcohol outlets also derived from mobile phone location data, and found a statistically significant and positive correlation between the two (Hu et al., 2021). Although the derived alcohol outlet visits cannot precisely represent the average level of alcohol use of different neighborhoods, they could be an effective predictor for enhancing DV prediction at the neighborhood level. In this work, we focus on the research question: *Are alcohol outlet visits derived from anonymized mobile phone location data an effective predictor for DV at the neighborhood level*? To answer this question, we conduct a series of experiments using four different statistical and machine learning models to understand the role of the derived alcohol outlet visits in DV prediction. Our research contributions are as follows:

- We propose an approach for deriving neighborhood-level alcohol outlet visits by integrating anonymized mobile phone location data, POI data, and census data.

- We investigate the role of the derived neighborhood-level alcohol outlet visits in enhancing DV prediction using different machine learning and statistical models.

Our study uses census block groups to approximate neighborhoods, which are the smallest geographic units with complete socioeconomic and demographic data from the American Community Survey (ACS). This enables DV prediction at a finer spatial scale than census tracts or zip code zones used in previous research (Beyer et al., 2015; McKinney et al., 2009; Voith & Brondino, 2017). The remainder of this paper is organized as follows. Section 2 reviews related work on individual-level and neighborhood-level DV studies. Section 3 describes the study area and data, and presents our methods for deriving alcohol outlet visits from mobile phone location data and the experiments for examining its role in enhancing DV prediction. Section 4 presents the experiment results, and Section 5 discusses the implications of the results. Finally, Section 6 concludes and summarizes this work.

## 2. Related Work

As a private form of violence, DV has often been examined at the individual level. Researchers generally focused on the variables related to the characteristics of individuals or couples, such as age, education, employment status, income, impulsivity, history of drug or alcohol use, and experience of childhood abuse, and investigated their associations with DV (Abramsky et al., 2011; Jewkes, 2002; Schafer et al., 2004). The methods used for collecting individual-level data were typically surveys and interviews. For example, Schafer et al. (2004) conducted face-to-face interviews with 1,635 couples living in the 48 contiguous US states to develop a path model of risk factors for intimate partner violence. Abramsky et al. (2011) studied the factors associated with intimate partner violence using a dataset of household surveys from the World Health Organization; the dataset was collected during a time period of 3 years via interviewing over 20,000 women. These surveys and interviews provide highly valuable individual-level insights about DV and its associated risk factors. Meanwhile, they cost substantial amounts of financial and human resources and often require a long time to complete.

There also exists a growing interest in understanding neighborhood-level factors that affect the occurrence of domestic violence. Researchers have examined some of the factors comparable to individual-level characteristics, such as the neighborhood-level average educational attainment, unemployment rate, and median household income (Beyer et al., 2015). Researchers also studied some other factors unique at the neighborhood level. Particularly, *Social Disorganization Theory* proposed by Shaw and Mckay (1942) has motivated the investigation of many neighborhood-level factors. For example, Voith and Brondino (2017) found that the concentrated disadvantage of a neighborhood and its residential instability are significantly related to the rate of intimate partner violence. They measured the concentrated disadvantage by a combination of variables, such as the proportion of households below poverty level, households on public assistance, and female-headed households with kids below age 18; they measured residential instability based on the proportion of renter-occupied housing units. Weir (2019) found that neighborhood-level income index, anti-social behaviors rate, the proportion of Black, Asian, and minority ethnic population, and population density are predictive of DV at the neighborhood level. The socioeconomic and demographic variables used in neighborhood-level studies are typically obtained from census surveys.

Existing studies have shown that alcohol use is strongly associated with domestic violence (Abramsky et al., 2011; Leonard & Quigley, 2017; O'Farrell et al., 2004). These studies were mostly conducted at the individual level, while neighborhood-level studies are much rarer. When alcohol is included as a risk factor in neighborhood-level studies, it is often represented by alcohol availability, such as alcohol outlet density (AOD) rather than actual alcohol use (McKinney et al., 2009; Livingston, 2011; McKinney et al., 2012; Mair et al., 2013; Waller et al., 2013). One possible reason is that AOD information is comparatively easier to obtain than alcohol use at the neighborhood level, since the former can be calculated based on the locations of alcohol outlets and predefined geographic boundaries, whereas the latter has to be obtained via representative surveys. The results of the studies based on AOD, however, are mixed: some studies found AOD predictive of DV rates (Cunradi et al., 2014; Freisthler et al., 2007; Livingston, 2010; Mannon, 1997; McKinney et al., 2009), while some others found no clear association between the two (Gorman et al., 1998; Waller et al., 2012). AOD is sensitive to the geographic units used for density calculation, and makes assumptions on the distances people travel to alcohol outlets (Gruenewald & Treno, 2000). The aggregation of alcohol outlets to geographic units was criticized for not differentiating outlet types (Gmel et al., 2016), although there exist studies that did examine different alcohol outlet types (e.g., liquor stores, nightclubs, and taverns) and related harms (Stockwell et al., 1992; Gruenewald et al., 1999; Livingston, 2011; Foster et al., 2017).

Our work differs from the existing literature in the following two aspects. First, we propose to derive alcohol outlet visits of the residents of different neighborhoods from anonymized mobile phone location data, and to use the derived visits as a proxy for neighborhood-level alcohol use. These alcohol outlet visits are derived by integrating anonymized mobile phone location data, POI data, and census data. Given the increasing availability of anonymized mobile phone location data and their large geographic coverage (e.g., the data used in this study cover the entire United States), we can efficiently derive neighborhood-level alcohol outlet visits for many geographic regions and therefore enable large geographic scale studies. Second, because the derived alcohol outlet visits cannot precisely represent alcohol purchase or alcohol use, we conduct a series of experiments to understand to what extent they can be used as a predictor for enhancing DV prediction at the neighborhood level. We utilize four different machine learning and statistical models to investigate the role of the derived alcohol outlet visits in DV prediction. These models also include a spatially explicit model that enables us to understand the spatially varied effect of alcohol outlet visits on DV across different geographic areas.

## 3. Methods

### 3.1. Study area and data

*Study area.* This study focuses on the City of Chicago, which is one of the major cities in the United States and the largest city in the US Midwest. We choose Chicago largely because of its data availability: the public data portal of Chicago provides access to its crime data that includes a column explicitly indicating whether an incident is DV. The geographic units of our study are census block groups (CBGs), and the time period of the data is from December 31, 2018 to January 6, 2020, which covers the entire year of 2019. We choose this time period because 2019 is the last year before the major disruption of COVID-19, and this time period also allows a potential follow-up study on the changes of DV rate and alcohol outlet visits during COVID-19. Figure 1(a) shows the city boundary of Chicago (dashed black outline) and its CBGs (blue polygons).

*Domestic violence data.* We downloaded the crime data from the Chicago Data Portal, which was originally from the Chicago Police Department's Citizen Law Enforcement Analysis and Reporting (CLEAR) System. This dataset contains all crimes reported by the police department and provides a number of important attributes about each crime incident, such as crime type (e.g., arson, burglary, and assault), location type (e.g., street and residence), latitude and longitude of the crime location, time of the crime, and others. As mentioned previously, this dataset contains a special column called "Domestic" with its values as either "True" or "False". According to its metadata, this column "indicates whether the incident was domestic-related as defined by the Illinois Domestic Violence Act". Thus, our initial attempt was extracting the incidents whose "Domestic" value is "True". However, after careful data examination, we noticed that a number of crime incidents, which are not about DV, such as arson, burglary, and theft, were mistakenly labeled as "True" in the column of "Domestic", possibly due to human errors when police entered the data. We filtered out those non-DV incidents. In addition, we noticed that the location types of some DV incidents are "Street" and other non-home locations, such as "Gas Station", "Movie Theater", and "Dental Office". Because our anonymized mobile phone location data can only link alcohol outlet visits to the home CBGs of the visitors, we removed those DV incidents that occurred in a public space and kept only those that occurred in home-related locations, such as "Residence", "Apartment", and "Driveway - Residential". Figure 1(b) shows the locations of the extracted DV incidents (red dots).

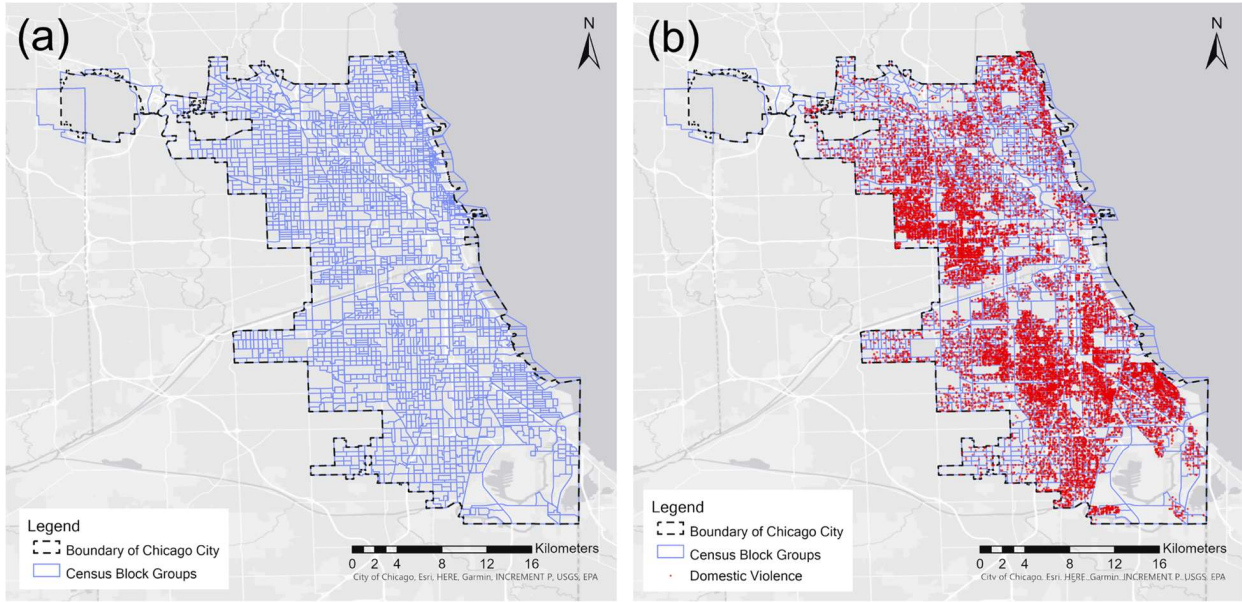

Figure 1. The study area and data: (a) Chicago city boundary and its CBGs; (b) locations of the extracted DV incidents.

*Anonymized mobile phone location data.* The anonymized mobile phone location data are provided by the company SafeGraph, which offered their data for the research community for free. Those location data were collected based on over 45 million anonymized smart mobile devices, most of which are smartphones, and locations were recorded by the GPS sensors in these smart devices. For privacy protection, the exact location coordinates of individuals are not available, and the data were aggregated to POIs and CBGs. SafeGraph provides over 3.6 million POIs covering the entire United States, and these POIs also include various types of alcohol outlets, such as liquor stores, bars, breweries, and wineries. Figure 2(a) illustrates this mobile phone location data with one POI (e.g., an alcohol outlet) and two CBGs. The data contain visits to this POI and the home CBGs of the POI visitors, and therefore we can know, e.g., the CBGs where the visitors of an alcohol outlet reside. The home CBG of a device was inferred by SafeGraph based on its nighttime locations in the previous six weeks. To protect the privacy of the mobile device users, the exact home locations of POI visitors are not available (we only know the general home CBG of a visitor) and the exact movement trajectories of people visiting these POIs are also not available.

A visit to a POI is identified based on the GPS point location of a mobile phone and the building footprint of the POI (e.g., the red dotted polygon in Fig. 2(a)), and only those visits whose durations are longer than 4 minutes are recorded in the data. The sizes of POIs vary depending on the actual building footprint of a POI. Note that the polygon building footprint data are not included in the default SafeGraph data, and that is why the polygon in Fig. 2(a) is depicted using a dotted line. SafeGraph also uses spatial clustering techniques to address potential issues related to GPS location recording (e.g., drifting GPS signals), and uses machine learning models with additional POI information (e.g., POI business types and opening hours) to determine the most likely visits under complex situations (e.g., when the GPS location point is close to multiple POIs that are located near each other). More details about the methods used by SafeGraph to identify POI visits are available in their technical guide (SafeGraph, 2021).

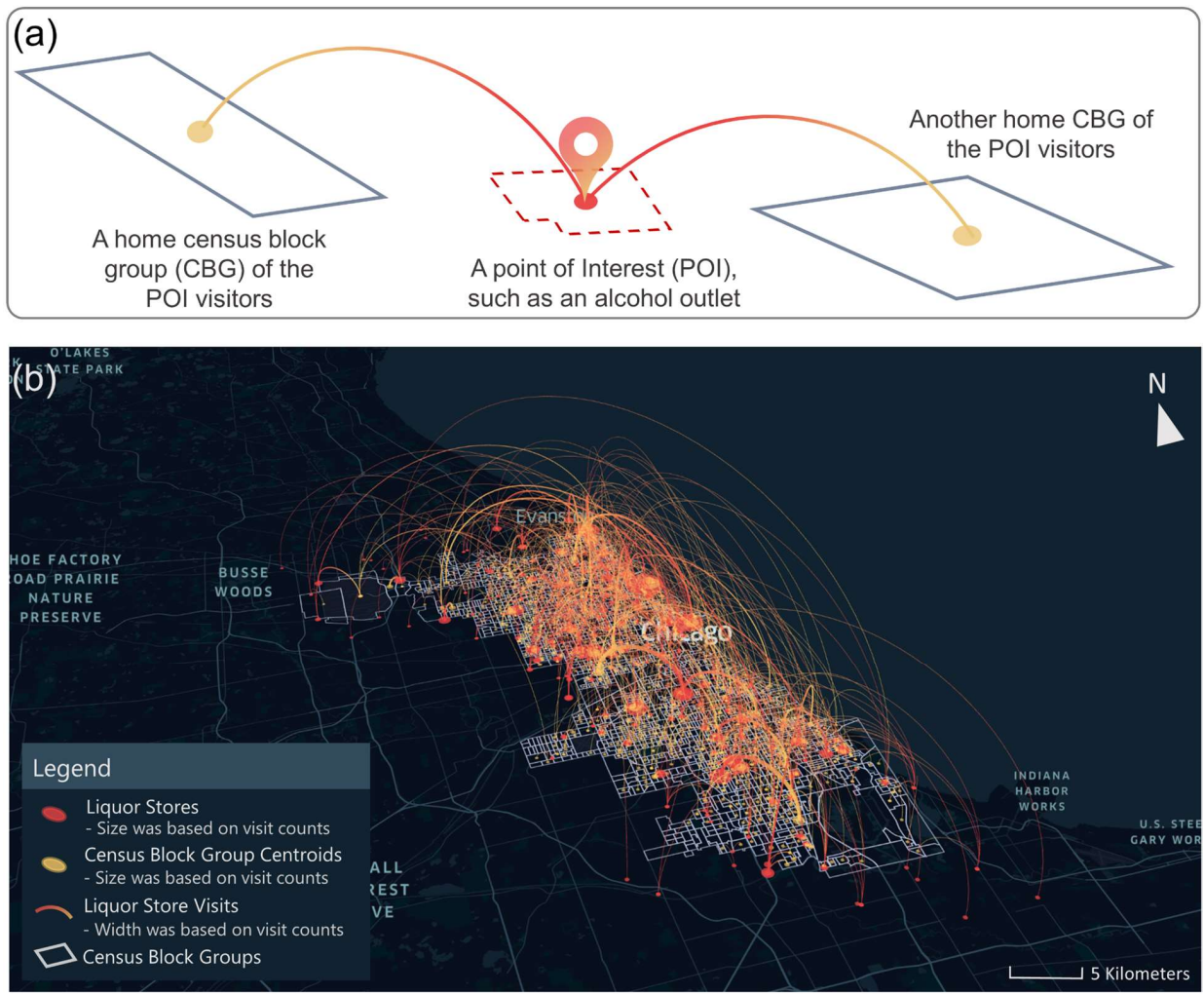

Figure 2. Anonymized mobile phone location data: (a) an illustration of the visits to a POI and two home CBGs of the POI visitors; (b) a map visualization of the visits to liquor stores in Chicago and the home CBGs of the liquor store visitors in a week of 2019.

By integrating mobile phone location data, POI data, and census CBGs, we can establish links between alcohol outlets and the home CBGs of outlet visitors based on their visiting behavior during a time period. Figure 2(b) provides a map visualization to illustrate people's visits from their home CBGs to liquor stores in Chicago in a week in 2019. In this map, each link represents the visits of people from their home CBG (represented as a polygon) to a liquor store (represented as a red filled circle). The thicker the link, the more visits from one CBG to one liquor store. To generate this map, we first fetched the liquor store POI data from the SafeGraph database. We then used the centroids of CBGs (represented as yellow filled circles) as the origins and the liquor store POIs as the destinations, and created links (visualized as arcs) among the origins and destinations. The map visualization in Fig. 2(b) was created using the open source geospatial tool, kepler.gl.

The anonymized mobile phone location data from SafeGraph were collected from over 45 million mobile devices, and this number is more than 10% of the total U.S. population (about 328 million according to the US Census). SafeGraph conducted a data quality evaluation by comparing its data with the ACS data from the US Census, and found that their data are statistically representative of the population at the county level and above (Squire, 2019). However, this data is not without limitations. The same analysis by SafeGraph also showed that their data may over- or under-represent the population at the CBG level (Squire, 2019). In addition, the data do not have socioeconomic and demographic information about the mobile device users for privacy protection reasons, and as a result, it is unclear whether the data can well represent different population groups (e.g., different age, ethnicity, and income groups). Coston et al. (2021) conducted an independent evaluation on SafeGraph data by comparing it with voter roll data, and they found that SafeGraph data, while capturing the overall human mobility pattern fairly well, underrepresent older (age over 65) and non-white population groups who are less likely to use smartphones. In this work, we do not assume that SafeGraph data can represent the population in different CBGs in Chicago, but use the data as a signal and examine its effectiveness for enhancing DV prediction.

## 3.2. Overview of experiment design

The goal of this study is to investigate whether the derived alcohol outlet visits can serve as an effective predictor for DV at the neighborhood level. Figure 3 provides an overview of the experiments designed for this investigation. We first derive alcohol outlet visit information from anonymized mobile phone location data and aggregate the derived information to the home CBGs of the alcohol outlet visitors. We derive visit information for four types of alcohol outlets: liquor stores, drinking places (pubs and bars), wineries, and breweries. Then, we conduct two sets of experiments to test the effectiveness of alcohol outlet visits for DV prediction. In the first set of experiments (baseline experiments), we predict DV rates at the neighborhood level using socioeconomic variables that are typically used in existing studies, and we do not include the derived alcohol outlet visits; in the second set of experiments (test experiments), we add the derived alcohol outlet visits in addition to the socioeconomic variables for DV prediction. We experiment with four types of machine learning and statistical models to examine the effectiveness of alcohol outlet visits as a predictor in different models. In the following, we present the methods for deriving measures and the models used for the experiments.

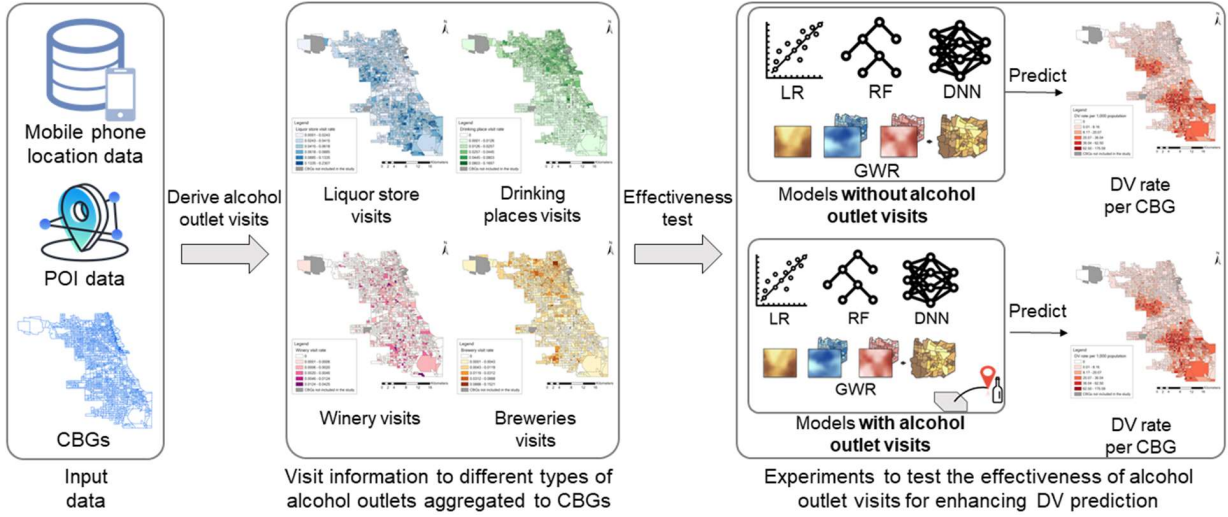

Figure 3. An overview of the experiments designed for testing the effectiveness of alcohol outlet visits for enhancing DV prediction at the neighborhood level.

### 3.3. Deriving domestic violence rate

The domestic violence rate at each CBG is the dependent variable that we aim to predict in this study. To obtain CBG-level DV rates, we first aggregate DV incidents to their corresponding CBGs based on their latitudes and longitudes using the spatial operation *within*, and compute the total number of DV incidents in each CBG. We then divide the total number of DV incidents in a CBG by the total population of this CBG (from census data), and multiply the result by 1,000 to obtain the DV incidents per 1,000 people during the period of December 31, 2018 to January 6, 2020. We apply the multiplication of 1,000 following the literature (Weir, 2019), and this operation also helps enlarge the obtained DV rate values and avoids potential underflow issues related to very small floating point numbers. Equations (1-2) summarize this computing process,

$$domestic\ violence\ rate\ for\ CBG_j = \frac{\sum_{i=1}^{m} DV_{ij}}{p_j} \times 1{,}000 \,, \qquad (1)$$

$$DV_{ij} = 1,\ if\ incident\ i\ is\ within\ CBG_j;\ DV_{ij} = 0,\ otherwise \,, \qquad (2)$$

where $DV_{ij}$ indicates whether incident $i$ is spatially within $CBG_j$, and $m$ is the total number of DV incidents in the data. $p_j$ is the total population of $CBG_j$ based on census data.

### 3.4. Deriving neighborhood-level alcohol outlet visit information

We propose to derive alcohol outlet visit information for the residents of different neighborhoods by combining anonymized mobile phone location data, POI data, and census CBG data. The first two types of data are from SafeGraph, and the CBG data are from the U.S. Census. We first extract alcohol outlets from the POI data using their North American Industry Classification System (NAICS) code. Specifically, we focus on four types of POIs that have a clear link to alcohol purchase: *Beer, Wine, and Liquor Stores* (445310), *Breweries* (312120), *Wineries* (312130), and *Drinking Places* (722410). The names of the four POI types are taken verbatim from the SafeGraph data, and the numbers in the parentheses are their corresponding NAICS codes. Note that *Drinking Places* here specifically refer to bars and pubs that sell alcoholic beverages. While these four types of alcohol outlets are included in this study, people can purchase alcohol from other types of places as well, such as *Grocery Stores* (445110), *Full-Service Restaurants* (722511), and *Limited-Service Restaurants* (722513). These other types of places are not included in our study, because the anonymized mobile phone location data do not contain information about the items people purchased at these places. In a recent study, we examined the capability of grocery store visits for predicting alcohol sales and found that grocery store visits are too noisy to provide improvement on alcohol sales prediction, and may even slightly decrease the performance of a predictive model (Hu et al., 2021). This is likely due to the fact that many visits to grocery stores do not involve alcohol purchase (although some grocery store visits do). Visits to the four types of alcohol outlets used in this study do not necessarily involve alcohol purchase either, but the likelihood that people purchase alcohol at these places are higher due to their business nature. Thus, we choose to focus on these four types of alcohol outlets in this study. Table 1 summarizes the numbers of outlets for each of the four types in the study area. Note that there are only 16 wineries, and this small number affects our analysis results later.

Table 1. The number of POIs in each of the four types of alcohol outlets.

| Alcohol outlet type | NAICS code | Number of POIs |
|---|---|---|
| *Breweries* | 312120 | 103 |
| *Wineries* | 312130 | 16 |
| *Beer, Wine, and Liquor Stores* | 445310 | 410 |
| *Drinking Places* | 722410 | 135 |

With the four types of alcohol outlets extracted from the POI data, we then utilize the anonymized mobile phone location data and census data to connect alcohol outlets to the home CBGs of the outlet visitors. The original SafeGraph data were organized with a focus on POIs, while providing information about the home CBGs of the POI visitors. Here, we reverse the focus of the data from POIs to CBGs, and compute the total number of visitors in each CBG who visited alcohol outlets. Figure 4 illustrates this process. Figure 4(a) shows the original organization of the SafeGraph mobile phone location data, in which the data focus on a POI (i.e., alcohol outlet 1 in the figure) and provides the home CBGs (i.e., CBG 1 to 4) of the alcohol outlet visitors to that POI. In other words, Fig. 4(a) shows that the visitors of alcohol outlet 1 reside in four different CBGs. Figure 4(b) shows our derived neighborhood-level alcohol outlet visits, which focus on CBGs (i.e., CBG 1) and provide the alcohol outlets that the residents of a CBG have visited (i.e., alcohol outlet 1 to 4). It is worth noting that the visited alcohol outlets may or may not be located within the CBG, e.g., alcohol outlet 1 and 2 are within CBG 1, whereas alcohol outlet 3 and 4 are outside of CBG 1. Through this reversal process, we obtain information about the number of residents in a CBG who have visited alcohol outlets in the city during a time period. Since these CBGs are the home neighborhoods of the alcohol outlet visitors, they are mostly residential areas rather than nighttime entertainment areas.

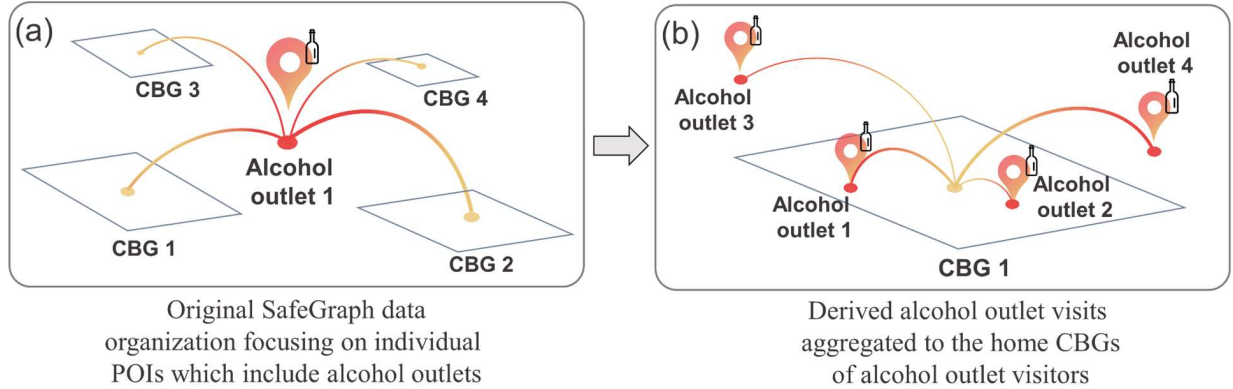

Figure 4. An illustration of deriving neighborhood-level alcohol outlet visits by integrating mobile phone location data, POI data, and CBG data.

We then obtain *alcohol outlet visit rate* by dividing the total number of visitors in a CBG using the total number of devices residing in the same CBG contained in the SafeGraph data. Such a normalization process is necessary since CBGs with higher population will naturally have larger numbers of alcohol outlet visitors than CBGs with lower population. Equation (3) shows the process of calculating alcohol outlet visit rate for a CBG,

$$alcohol\ outlet\ visit\ rate\ for\ CBG_j\ =\ \frac{\sum_{i=1}^{n} v_{ij}}{S_j}\ ,\qquad(3)$$

where $v_{ij}$ is the number of visitors in $CBG_j$ to an alcohol-related $POI_i$; $n$ is the total number of alcohol outlets in one type (e.g., liquor store) in the study area; $S_j$ is the total number of mobile devices residing in $CBG_j$ (inferred by SafeGraph based on devices' nighttime locations in the past six weeks). We calculate the visit rate for each CBG and for each of the four types of alcohol outlets. As a result, we obtain four alcohol outlet visit rates for each CBG in Chicago.

### 3.5. Socioeconomic and demographic variables

Previous studies have shown that a variety of socioeconomic and demographic variables at the neighborhood level can be utilized to predict DV (Beyer et al., 2015; Cunradi et al., 2011; Gorman et al., 1998; McKinney et al., 2009; Voith & Brondino, 2017; Waller et al., 2012; Weir, 2019). These neighborhood-level variables are often motivated by theories and frameworks, such as *Social Disorganization Theory* (Pinchevsky & Wright, 2012; Shaw & McKay, 1942) and the *Social Determinants of Health* framework (Hampton et al., 2003; Vives-Cases et al., 2011). Learning from the literature, we select neighborhood-level socioeconomic and demographic variables in five categories, which are: (1) race and ethnicity, (2) age, (3) social disadvantage level,

(4) residential instability, and (5) urbanicity. We then add the derived neighborhood-level alcohol outlet visits as the 6th category of variables, and investigate their effectiveness for enhancing DV prediction at the neighborhood level. Table 2 summarizes these independent variables and provides their descriptions.

Table 2. Notations and descriptions of the independent variables.

| Variable notations | Descriptions |
| --- | --- |
| **(1) Race and ethnicity** | |
| *% White* | Percentage of population in White |
| *% Black* | Percentage of population in Black or African American |
| *% Ame Indi and AK Native* | Percentage of population in American Indian and Alaska Native |
| *% Asian* | Percentage of population in Asian |
| *% Nati Hawa and Paci Island* | Percentage of population in Native Hawaiian and Other Pacific Islander |
| *% other races* | Percentage of population in some other race alone, or two or more races |
| *% Hispanic* | Percentage of Hispanic or Latino population |
| **(2) Age** | |
| *% age <18* | Percentage of population below age 18 |
| *% age 18-29* | Percentage of population between age 18 to 29 |
| *% age 30-39* | Percentage of population between age 30 to 39 |
| *% age 40-49* | Percentage of population between age 40 to 49 |
| *% age 50-59* | Percentage of population between age 50 to 59 |
| *% age >60* | Percentage of population over age 60 |
| **(3) Social disadvantage level** | |
| *med income* | Median household income |
| *% unemployment* | Percentage of unemployed labor force population |
| *% female hh* | Percentage of households with children below age 18 headed by female |
| *% <highschool* | Percentage of population age 25 or over without high school completion |
| *% >highschool* | Percentage of population age 25 or over with high school completion and without bachelor's degree |
| *% >university* | Percentage of population age 25 or over with bachelor's degree or higher degree |
| *% security inc* | Percentage of households received supplemental security income in the past 12 months |
| *% assistant inc* | Percentage of households received public assistance income in the past 12 months |
| *% assist inc or snap* | Percentage of households received public assistance income or Food Stamps/ supplemental nutrition assistance program in the past 12 months |
| **(4) Residential instability** | |
| *% renter hh* | Percentage of renter-occupied housing units |
| *% stay >=5yrs* | Percentage of occupied housing units that householder has stayed for |

| | 5 years or more |
|---|---|
| **(5) Urbanicity** | |
| *population density* | Population density as people per square kilometer |
| **(6) Alcohol outlet visits** | |
| *brewery vr* | Brewery visit rate |
| *drinking place vr* | Drinking place visit rate |
| *liquor store vr* | Liquor store visit rate |
| *winery vr* | Winery visit rate |

All the socioeconomic and demographic variables in the five categories are directly obtained or derived from the ACS data of the US Census. For (1) race and ethnicity composition at the CBG level, we calculate the percentage of populations in different racial and ethnic groups. Similarly, we calculate the percentage of populations in each age group to measure (2) age composition of CBGs. For (3) the social disadvantage levels of the CBGs, we select variables including median household income, percentage of unemployed labor force population, percentage of households with children headed by female, percentage of population by educational attainment status, and percentage of households receiving different types of public assistance (e.g. public assistance income and supplemental nutrition assistance) in each CBG. For (4) residential instability, we use two variables, namely percentage of renter-occupied housing units and percentage of occupied housing units that householders have stayed in for 5 years or more. Finally, for (5) urbanicity, we compute the population density of each CBG by dividing its population using its total area from the TIGER data of the US Census.

### 3.6. Statistical and machine learning models

We employ four different statistical and machine learning models to examine the effectiveness of the derived alcohol outlet visits for enhancing DV prediction at the neighborhood level. These four models are: ordinary least squares (OLS), random forest (RF), deep neural network (DNN), and geographically weighted regression (GWR). Among them, OLS and GWR are statistical models while RF and DNN are machine learning models. We choose to use four different models, instead of only one, for three major reasons. First, there has been an increasing interest in using artificial intelligence (AI) and machine learning models for health research (Boulos et al., 2019). Therefore, including machine learning models in addition to statistical models in this study can help us understand whether the derived alcohol outlet visits are also effective in those machine learning models. Results obtained can also help establish baselines for other researchers who would like to explore the use of machine learning models in this direction. Second, the four models have different model architectures and take into account different factors, e.g., GWR considers spatial autocorrelation while OLS does not. The training process of both RF and DNN involves some degree of randomness, such as the random subsets of data used for training RF and the randomness in parameter initiation and optimization for training DNN (Géron, 2019). These different model architectures and randomness suggest that the effectiveness of alcohol outlet visits in one particular model does not necessarily guarantee their effectiveness in another model. Using four different models allows us to test their effectiveness in a more robust manner. Third, these different models offer complementary insights. OLS and RF can reveal the overall relative importances of the independent variables via regression coefficients and feature importance; DNN enables the examination of alcohol outlet visits in a deep learning model; and GWR allows an exploration of the spatially varying effects of alcohol outlet visits on DV prediction. Below, we provide some general methodological details of the four models.

- *Ordinary Least Squares*: OLS models the relation between multiple input features and the target variable via a linear equation. Specifically, the OLS model used in this work is in the form of Equation (4),

$$dv\ rate\ =\ \theta_0\ +\ \theta_r r\ +\ \theta_a a\ +\ \theta_s s\ +\ \theta_e e\ +\ \theta_u u\ (+\theta_l l)\ +\ \varepsilon\ , \tag{4}$$

where $\theta_r, \theta_a, \theta_s, \theta_e, \theta_u$ are the coefficients for the five categories of socioeconomic and demographic variables (i.e., race, age, social disadvantage, residential instability, and urbanicity) respectively, and $\theta_l$ are the coefficients for the four types of alcohol outlet visits. $\theta_l l$ is put within a pair of parentheses in Equation (4), because alcohol outlet visits will not be included in the first set of experiments that use socioeconomic and demographic variables only. Each of $\theta_r, \theta_a, \theta_s, \theta_e, \theta_l$ contains multiple coefficients for the variables in that category (e.g., $\theta_l$ contains four coefficients for the four types of alcohol outlet visits). It is worth noting that OLS is generally not suitable for data with spatial autocorrelation, which can bias the estimated coefficients. Here, we use OLS as a preliminary model for exploratory analysis, and it also provides a baseline to be compared with the GWR model that takes into account spatial autocorrelation. We use Python and the *statsmodels* library to implement the OLS model.

- *Random Forest*: RF is an ensemble machine learning model that consists of many decision trees trained on randomly selected subsets of the training data. The RF model then computes the average of the predictions from the many constructed decision trees. Compared with the OLS model that assumes a linear relation between the input features and the target variable, RF can model any complex nonlinear relation between the variables, which makes RF a powerful predictive model (Xia et al., 2021). We use Python and the *scikit-learn* library to implement the RF model. We perform hyperparameter tuning to identify the optimal parameters for the model. Following the literature (Liaw & Wiener, 2002; Huang et al., 2021), we focus on tuning two major hyperparameters, namely $n_{tree}$, which controls the number of decision trees, and $m_{try}$, which controls the number of features to consider at each split in a tree. We perform grid search to identify the best hyperparameters, and the search space for $n_{tree}$ is set to [10, 200] with an interval of 10, and the search space for $m_{try}$ is $\{S, \sqrt{S}, log_2 S, S/2, S/3\}$ where $S$ is the number of input features. Based on the result of hyperparameter tuning, we set $n_{tree}$ as 80 and $m_{try}$ as $\sqrt{S}$ for the RF model.

- *Deep Neural Network*: DNNs and other deep learning models have shown outstanding prediction capabilities in recent years (LeCun et al., 2015; Hu et al., 2019). A DNN uses multiple layers of neurons to learn a complex nonlinear relation between the input features and the target variable. The architecture of a DNN model is highly flexible, and one can choose different numbers of layers and different numbers of neurons per layer. Other techniques, such as drop out, can also be applied to each layer to reduce overfitting and improve the performance of the model. We use Python and the *TensorFlow* library to implement a DNN model. We perform hyperparameter tuning to identify an optimal model architecture for the DNN model following the literature (Géron, 2019). Specifically, we perform random search to determine the number of hidden layers and number of neurons per layer. The search space for the number of layers is [1, 10], and for each layer, the search space for neurons is [16, 256] with an interval of 16. We also test the performance of the model with dropout rates added or not added to each layer. Based on the hyperparameter tuning result, we eventually build a DNN model that has four hidden layers with 128, 128, 64, and 32 neurons respectively. A dropout rate of 0.2 is added to the first two hidden layers. The ReLU activation function is used in all layers.

- *Geographically Weighted Regression*: GWR is a spatially explicit model that fits local OLS models for each geographic unit (i.e., CBG in this study) by taking into account spatial autocorrelation and spatial heterogeneity (Brunsdon et al., 1998). Specifically, the GWR model used in this work is in the form of Equation (5),

$$dv\ rate\ =\ \theta_0(x_i, y_i)\ +\ \theta_r(x_i, y_i)\ r\ +\ \theta_a(x_i, y_i)\ a\ +\ \theta_s(x_i, y_i)\ s\ +$$
$$\theta_e(x_i, y_i)\ e\ +\ \theta_u(x_i, y_i)\ u\ (+\theta_l(x_i, y_i)\ l)\ +\ \varepsilon_i\ , \tag{5}$$

where $(x_i, y_i)$ is the spatial coordinates of a geographic unit (i.e., a CBG in this study). The coefficients have the same meaning as used in the OLS model, but are functions of spatial coordinates, i.e., the coefficients will vary across different geographic locations capturing the spatial effects between the independent and target variables. To configure the model, we employ the typically used Gaussian kernel and apply a golden search approach to identify the optimal bandwidth by minimizing the AIC value (Oshan et al., 2019). The identified optimal bandwidth is 243 nearest CBGs, which is then used in the experiments. We use Python and the *mgwr* library (Oshan et al., 2019) to implement the GWR model.

With these four different models, we conduct two sets of experiments: the baseline experiments use only socioeconomic and demographic variables (i.e., the variables in categories (1) to (5) in Table 2), and the test experiments use alcohol outlet visits in addition to the socioeconomic and demographic variables (i.e., the variables in categories (1) to (6)). Two goodness-of-fit measures, $R^2$ and root mean square error (RMSE), are utilized for assessing the performance of the four models. For the statistical models, namely OLS and GWR, their $R^2$ and RMSE can be directly obtained from the model fitting results. For the machine learning models, namely RF and DNN, their $R^2$ and RMSE are obtained via a 10-fold cross-validation process, in which the data are divided into 10 non-overlapping folds, and the training and test procedure are iterated 10 times. In each iteration, one of the 10-fold data is held out as the test data, and the remaining 9-fold data are used for training the model. The mean $R^2$ of the ten iterations is reported, and the RMSE is calculated by pooling the prediction residuals from the ten iterations. Note that the same random seed is used in the two sets of experiments to separate the data into the exactly same 10 folds. Such a 10-fold cross-validation process can help obtain more robust evaluation results and avoid potential biases caused by one particular validation dataset.

In addition to $R^2$ and RMSE, we also report adjusted $R^2$ and Akaike information criterion (AIC) for OLS and GWR, which take into account model complexity. Adjusted $R^2$ and AIC can tell us whether the improved prediction (if any) of the model is due to the useful information provided by the newly added variables (e.g., neighborhood-level alcohol outlet visits in this study), or whether the improvement is simply due to the increased complexity of the model. In total, we have four evaluation measures, namely $R^2$, RMSE, adjusted $R^2$, and AIC. These evaluation measures are selected based on the types of models used in this study. Additional measures, such as Widely Applicable Information Criterion (WAIC) (Watanabe & Opper, 2010), could also be employed when other types of models (e.g., Bayesian models) are used.

## 4. Results

### 4.1. Derived neighborhood-level DV rates and alcohol outlet visit rates

The derived neighborhood-level DV rates are visualized in Fig. 5(a), where one can observe that neighborhoods with high DV rates tend to be clustered in the western and southern areas of the city. Given this observation, we compute the global Moran's $I$ index (Fig. 5(b)) to examine the spatial autocorrelation of neighborhood-level DV rates. We find that the DV rate in Chicago has a statistically significant and positive spatial autocorrelation, with a Moran's $I$ index 0.648 ($p<0.001$), which suggests that neighborhoods with high DV rates tend to be clustered with each other in Chicago, rather than being distributed more evenly in the city.

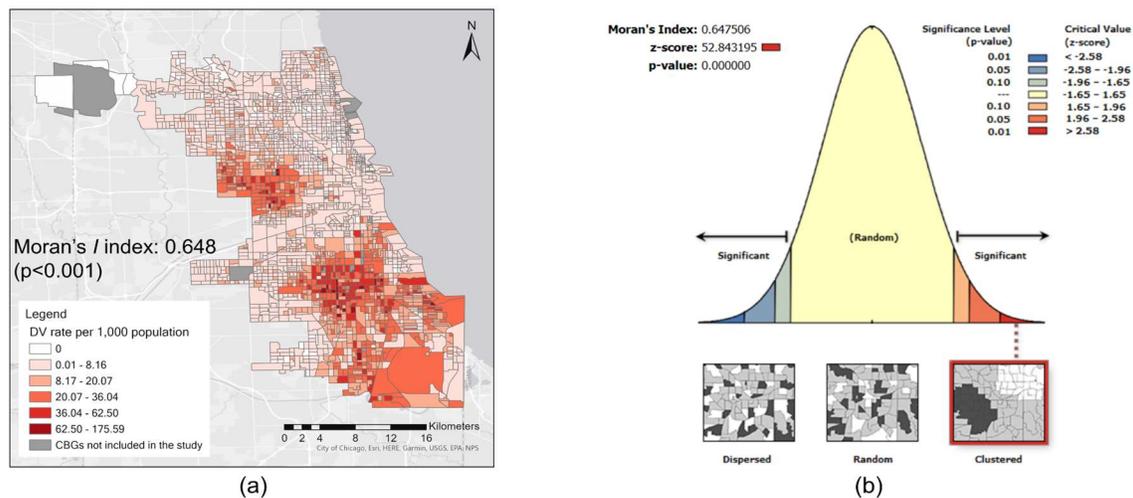

(a)                                                                    (b)

Figure 5. Neighborhood-level DV rates in Chicago and global Moran's $I$ test: (a) DV rates of the CBGs; (b) Moran's $I$ test result.

In Figure 6, we visualize the neighborhood-level alcohol outlet visit rates derived from anonymized mobile phone location data by showing the visit rates for (a) liquor stores, (b) drinking places, (c) breweries, and (d) wineries respectively.

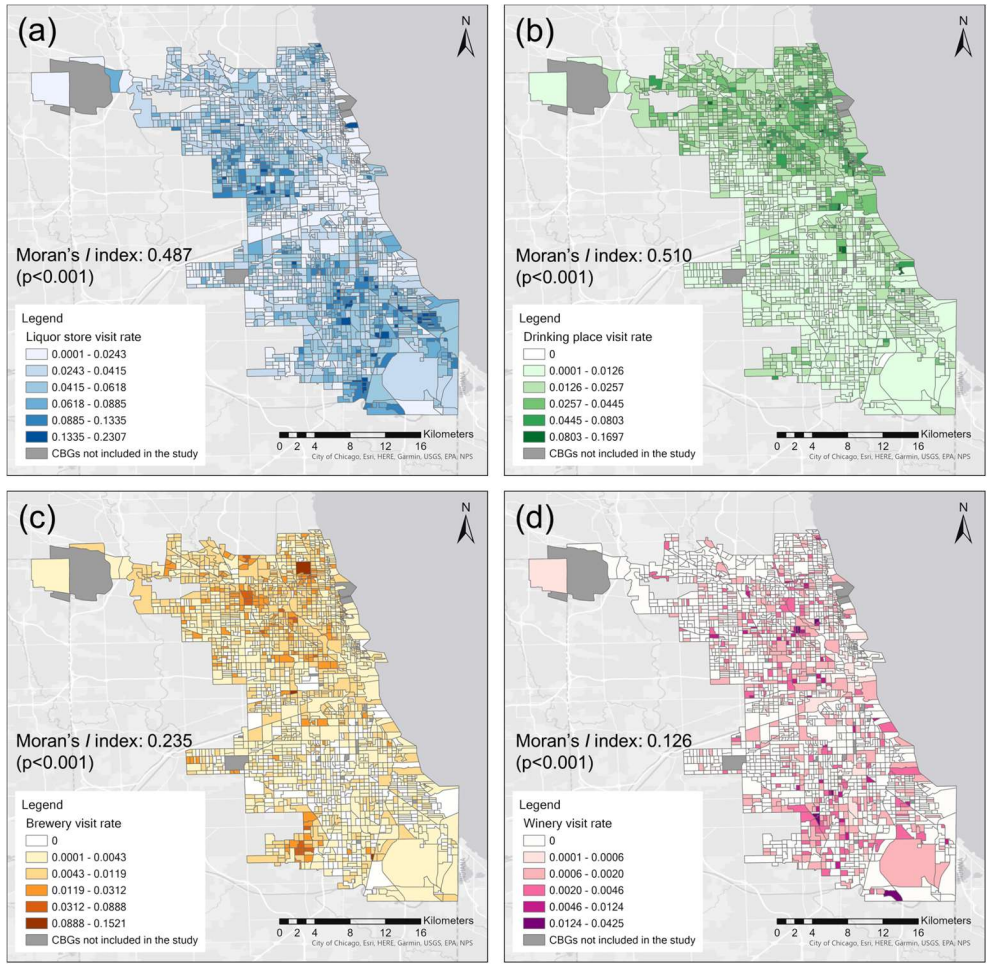

Figure 6. Neighborhood-level alcohol outlet visit rates: (a) liquor stores; (b) drinking places; (c) breweries; (d) wineries.

Similar to DV rates, we also examine the spatial autocorrelation of the derived alcohol outlet visits. We perform global Moran's $I$ tests, and the obtained Moran's $I$ indices are shown in the subfigures of Fig. 6. The Moran's $I$ indices are all positive and statistically significant ($p<0.001$), suggesting that there exist statistically significant spatial autocorrelations in the visit rates of all four types of alcohol outlets at the CBG level. This means that CBGs with similar alcohol outlet visit rates tend to cluster with each other, rather than being distributed more evenly in the city. In addition, the Moran's $I$ indices of liquor store visits and drinking place visits are about twice to four times as large as the indices of brewery visits and winery visits. This result suggests that the spatial autocorrelations of the visits to liquor stores and drinking places are stronger than those of the visits to the other two types of alcohol outlets. Looking at subfigure (a) for liquor store visit rates, we can see two major clusters of CBGs with high liquor store visit rates located in the western and the southern areas of the city; looking at subfigure (b) for drinking place visit rates, we can see one major cluster of CBGs with high drinking place visit rates in the northern area of the city. In addition to map visualizations, we also plot out the histograms of the four types of alcohol outlet visit rates in the supplementary Fig. S1. We observe a small number of CBGs in each histogram with high outlet visit rates, although the frequency distributions of the visit rates to liquor stores and drinking places are less skewed than the other two types of outlets.

## 4.2. Correlation analysis

We explore the general relationship between the rates of DV and those of the four types of alcohol outlets by performing correlation analysis. Considering that the relations may not be linear, we perform both Pearson's correlation and Spearman's correlation analyses, and the results are reported in Table 3. The two types of correlation analyses show consistent results: liquor store visit rates have a statistically significant and positive correlation with DV rates, while visit rates to drinking places and breweries show statistically significant but negative correlation with DV rates. The correlation between DV rates and winery visit rates is not significant, probably due to the small number of wineries in the study area.

Table 3. Correlation coefficients and their P-values between the four types of alcohol outlet visit rates and DV rates.

| | Pearson's correlation | | Spearman's correlation | |
|---|---|---|---|---|
| | Coefficient | P-value | Coefficient | P-value |
| liquor store vr | 0.422*** | 0.000 | 0.505*** | 0.000 |
| drinking place vr | -0.342*** | 0.000 | -0.493*** | 0.000 |
| brewery vr | -0.197*** | 0.000 | -0.384*** | 0.000 |
| winery vr | 0.026 | 0.233 | 0.040 | 0.067 |

*p-value < 0.05; **p-value < 0.01; ***p-value < 0.001.

## 4.3. Multicollinearity diagnosis

Before examining the role of alcohol outlet visits for DV prediction in statistical and machine learning models, we carry out a series of diagnostic tests to check multicollinearity among the independent variables. To do so, we compute the variance inflation factor (VIF) for the socioeconomic and demographic variables. It is worth noting that VIF assumes that the analyzed data meet the assumptions of OLS, which may not hold for spatial data. Nevertheless, VIF can still help identify multicollinearity to some extent, and it was used in existing studies on spatial data (Castaldi et al., 2016; Li & Lam, 2018). Therefore, it is used here to help select a subset of variables in order to reduce the multicollinearity among the initial set of variables. We standardize all variables using their mean and standard deviation, and then compute the VIF values. With the obtained VIF values, we gradually remove the variables with very high VIF values until they are all smaller than the typical cut-off threshold 5. Table 4 shows the results of these VIF tests.

Table 4. VIF values obtained from the multicollinearity tests.

| Variable | First test | Second test | Final test |
|---|---|---|---|
| % White | > 1000.000 | 16.399 | 2.965 |
| % Black | > 1000.000 | 29.665 | - |
| % Ame Indi and AK Native | > 1000.000 | 1.072 | 1.037 |
| % Asian | > 1000.000 | 2.979 | 1.216 |
| % Nati Hawa and Paci Island | > 1000.000 | 1.010 | 1.010 |
| % other races | > 1000.000 | - | - |
| % Hispanic | 7.417 | 7.408 | 3.219 |
| % age <18 | > 1000.000 | - | - |
| % age 18-29 | > 1000.000 | 2.931 | 2.730 |
| % age 30-39 | > 1000.000 | 2.955 | 2.793 |
| % age 40-49 | > 1000.000 | 1.785 | 1.726 |
| % age 50-59 | > 1000.000 | 1.690 | 1.658 |
| % age >60 | > 1000.000 | 3.064 | 2.983 |
| med income | 3.675 | 3.672 | 3.008 |
| % unemployment | 1.844 | 1.842 | 1.733 |
| % female hh | 2.830 | 2.827 | 2.668 |
| % <highschool | > 1000.000 | 3.118 | 2.847 |
| % >highschool | > 1000.000 | - | - |
| % >university | > 1000.000 | 6.238 | - |
| % security inc | 1.804 | 1.798 | 1.628 |
| % assistant inc | 1.436 | 1.433 | 1.265 |
| % assist inc or snap | 4.998 | 4.993 | - |
| % renter hh | 2.949 | 2.948 | 2.692 |
| % stay >=5yrs | 2.370 | 2.366 | 2.331 |
| population density | 1.176 | 1.176 | 1.146 |

As can be seen, the first test shows that many variables have extremely high VIF values, suggesting severe multicollinearity for these variables. This is likely due to the fact that the variables in three categories (i.e., race, age, and educational attainment) are composition measures that add up to 1. Accordingly, in the second test, we remove one variable from each of the three categories, which are "% other races", "% age <18", and "% >highschool". The result of the second test shows substantially reduced VIF values but some of these values are still larger than 5. In the final test, we further remove three variables with high VIF values in two categories, which are "% Black" in the race category and "% >university" and "% assist inc or snap" in the social disadvantage category. Note that the VIF value of "% assist inc or snap" is 4.993 in the second test. While it is not strictly larger than the cut-off value 5, it is at the threshold boundary and partially overlaps with another variable "% assistant inc". Therefore, we also remove this variable. After these three variables are removed, the final test shows that the VIF values of all remaining variables are smaller than 3.3, suggesting low multicollinearity among them. We therefore use these remaining 19 socioeconomic and demographic variables in the experiments.

### 4.4. Results from the four statistical and machine learning models

*Ordinary least squares.* We first test the effectiveness of the derived alcohol outlet visits for enhancing DV prediction at the neighborhood level using the OLS model. In the baseline experiment, OLS uses only the 19 socioeconomic and demographic variables as the independent variables; whereas in the test experiment, OLS uses the four alcohol outlet visit variables (i.e., visit rates for the four types of alcohol outlets) in addition to the 19 socioeconomic and demographic variables (23 variables in total). In the baseline experiment, we obtain an $R^2$ of 0.528 and an RMSE of 9.653, whereas in the test experiment, we obtain an increased $R^2$ of 0.548 and a decreased RMSE of 9.452. Considering that the model improvement may be simply due to the inclusion of more independent variables, we further compute the adjusted $R^2$ and AIC values. We find that the adjusted $R^2$ has increased from 0.524 to 0.543, and the AIC has decreased from 15861.358 to 15778.896. This result suggests that the derived alcohol outlet visits indeed provide useful information that helps improve the prediction of the OLS model, even when the increased model complexity is taken into account.

We then examine the regression coefficients and their *p* values obtained via the OLS model, which are summarized in Table 5. Consistent with the literature (Beyer et al., 2015; Weir, 2019), several variables related to poverty (e.g., *med income* and *% unemployment*), low educational attainment (*% < highschool*), and residential instability (*% renter hh*) show statistically significant relation with the DV rate of the corresponding CBG. High percentages of population in the groups of 50-59 and above 60 also show significant relation with the DV rate, which may be worth further investigation. Our analysis also shows several racial and ethnical groups (i.e., % White, % Asian, and % Hispanic) have statistically significant relations with the DV rate. These racial and ethnic variables, however, are likely to be intertwined with the socioeconomic status of these population groups in Chicago. Among the four alcohol outlet visit variables, three (except visits to wineries) have statistically significant relations with DV rate. This result is consistent with the correlation analysis, and the insignificant correlation between winery visits and DV rate is likely affected by the fact that there are only 16 wineries in the study area.

Table 5. Coefficients and their p-values obtained by the OLS model.

| Variable | Coefficient | P-value |
| --- | --- | --- |
| % White | -3.9003*** | 0.0000 |
| % Ame Indi and AK Native | 0.0649 | 0.7567 |
| % Asian | -2.3958*** | 0.0000 |
| % Nati Hawa and Paci Island | -0.2924 | 0.1567 |
| % Hispanic | -3.2277*** | 0.0000 |
| % age 18-29 | 0.0126 | 0.9706 |
| % age 30-39 | 0.3962 | 0.2530 |
| % age 40-49 | 0.4595 | 0.0902 |
| % age 50-59 | 0.5299* | 0.0454 |
| % age >60 | 0.8264* | 0.0207 |
| med income | -0.7119* | 0.0470 |
| % unemployment | 1.8048*** | 0.0000 |
| % female hh | 0.3272 | 0.3304 |
| % <highschool | 1.8482*** | 0.0000 |
| % security inc | -0.1241 | 0.6372 |
| % assistant inc | 0.0574 | 0.8039 |
| % renter hh | 1.6061*** | 0.0000 |
| % stay >=5yrs | -0.6017 | 0.0551 |
| population density | -0.8653*** | 0.0001 |
| brewery vr | -0.5707** | 0.0080 |
| drinking place vr | -0.9868*** | 0.0001 |
| liquor store vr | 1.9857*** | 0.0000 |
| winery vr | -0.2995 | 0.1485 |

*p-value < 0.05; **p-value < 0.01; ***p-value < 0.001.

Since all independent variables are standardized, we can compare their coefficients to understand their relative impacts on the DV rate at the CBG level. Figure 7 shows these significant variables ranked based on their coefficients from large to small. Note that we also include the coefficient of *winery vr* in the figure (although its relation with DV is insignificant), since it is one of the main variables examined here.

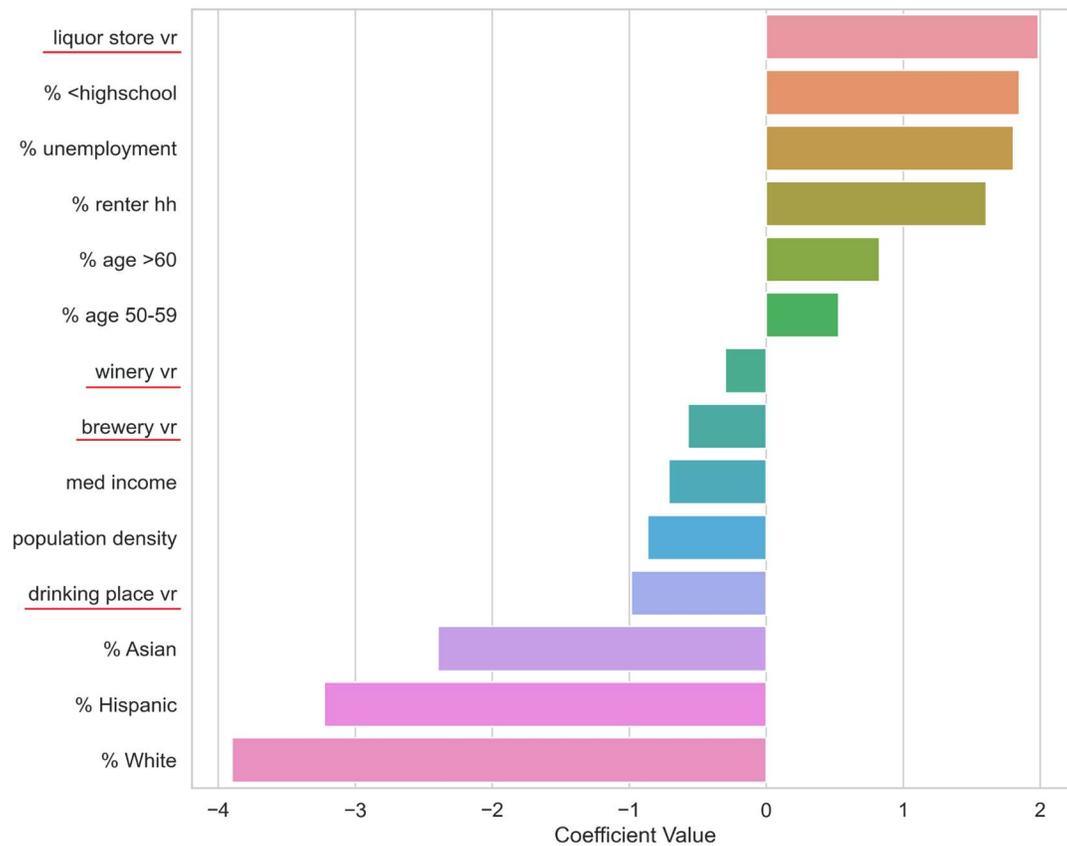

Figure 7. Coefficients of the significant variables from the OLS model in a descending order.

Remarkably, the variable of liquor store visit rates shows the highest positive contribution to DV rate compared with other variables. This result suggests that liquor store visitation is an important predictor for DV rate at the neighborhood level. Interestingly, visits to two other types of alcohol outlets (i.e., drinking places and breweries) have negative coefficients, which are consistent with the correlation analysis. This result indicates that visits to these outlets are, in fact, associated with lower DV rates. This result is surprising, as a higher density of drinking places in a neighborhood was often thought to be related to a higher DV rate (Cunradi et al., 2011; Waller et al., 2012). On the other hand, higher visit rates to drinking places and breweries could also imply better financial capabilities and higher socioeconomic statuses of the visitors who can afford going to these places, which could be linked to lower DV rates.

*Random forest.* The RF model follows the same experiment setting as used for the OLS model: the baseline experiment uses the 19 socioeconomic and demographic variables selected through the multicollinearity tests, and the test experiment uses the derived four alcohol outlet visit rate variables in addition to the 19 socioeconomic and demographic variables. The RF model achieves an $R^2$ of 0.594 and an RMSE of 8.978 in the baseline experiment, and an increased $R^2$ of 0.612 and a decreased RMSE of 8.848. This result indicates that the derived alcohol outlet visits can help the RF model improve its DV prediction as well.

The output of the RF model includes *feature importance*, which allows us to examine the relative importance of different input variables (or *features,* as typically called in machine learning literature) for DV rate prediction. The feature importance values output by a RF model are in the range of [0, 1], and there is one importance value for each input feature; together, these importance values sum up to 1. An importance value suggests the relative capability of a feature in helping the RF model make correct predictions. Thus, a feature with a higher importance value is more important than another feature with a lower importance value for the RF model to make correct predictions. One caution, however, we need to take when interpreting the results of machine learning models is that the best predictors may not be theoretically informative. Figure 8 shows the feature importance of all independent variables, including the four alcohol outlet visit variables. Since we have used 10-fold cross-validation, ten RF models are trained which result in ten sets of feature importance. Therefore, we use box plots to represent them in Fig. 8.

Among the 23 variables, the top six important predictors are related to the demographic and socioeconomic status of the neighborhoods, including *% White*, *med income*, *% unemployment*, *% Hispanic*, *% Asian*, and *% female hh*. Note that unlike the coefficients obtained in the OLS model, feature importance only tells us the relative strength of the contribution made by an input variable in helping the RF model make correct predictions, and does not provide information on whether a variable is positively or negatively associated with DV. Visits to liquor stores are ranked as the 7th most important variable for DV prediction at the neighborhood level, and visits to drinking places and to breweries are ranked as the 9th and 16th important among the 23 input variables. Similar to the result of the OLS model, winery visit rates seem to be less useful for DV prediction in this study.

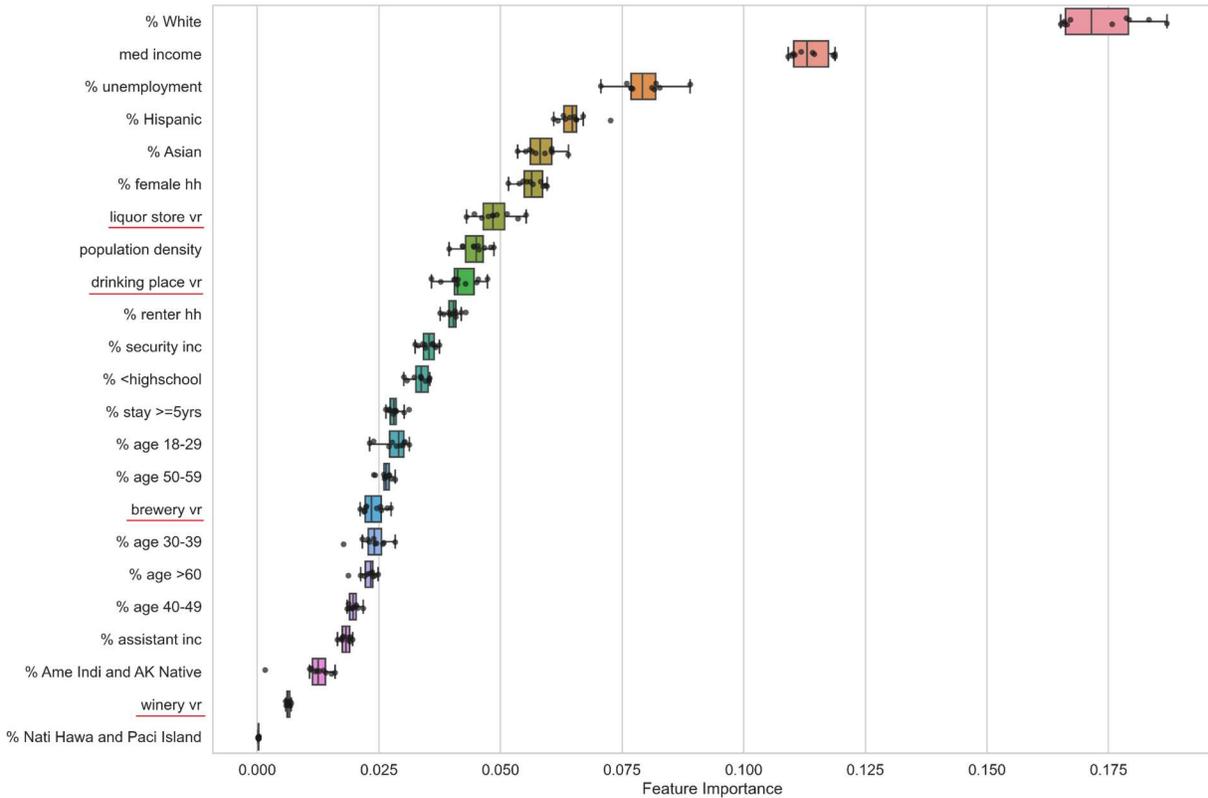

Figure 8. Feature importance of the input variables for the RF model obtained through the 10-fold cross-validation process.

*Deep neural network.* For the DNN model, the baseline experiment with only socioeconomic and demographic variables achieves an $R^2$ of 0.575 and an RMSE of 9.250. After the inclusion of the four alcohol outlet visit variables, the $R^2$ of the model increases to 0.595 and the RMSE decreases to 9.018. This result shows that the derived neighborhood-level alcohol outlet visits are also effective for the DNN model to improve its DV prediction. While DNNs are powerful predictive models, they function like a "black box", and consequently, we cannot derive much additional information from the result (such as feature importance from the RF model). However, it does demonstrate the effectiveness of the derived alcohol outlet visits for DV prediction in a deep learning model.

*Geographically weighted regression.* For GWR, the baseline experiment using only socioeconomic and demographic variables achieves an $R^2$ of 0.774 and an RMSE of 6.681. After the inclusion of the four alcohol outlet visit variables, the $R^2$ increases to 0.796 and the RMSE reduces to 6.352. Similar to the OLS model, we compute the adjusted $R^2$ and AIC values of the models, so that the increased model complexity is taken into account. The adjusted $R^2$ increases from 0.721 in the baseline experiment to 0.737 in the test experiment when the alcohol outlet visit variables are added, and the AIC decreases from 15056.949 to 14987.023. This result shows that the derived alcohol outlet visits are again effective in a GWR model for enhancing DV prediction at the neighborhood level. In addition, since the GWR model produces local coefficients, we can further analyze them to understand the changing role of alcohol outlet visits across different locations. We provide a more detailed discussion in Section 5.

*Summary.* Table 6 provides a summary of the obtained experiment results. Overall, these results show that the alcohol outlet visits derived from anonymized mobile phone location data using our proposed method effectively improve the prediction of DV rate at the neighborhood level. As shown in the table, after alcohol outlet visits have been added into the input variables, the $R^2$ of all models increase while their RMSE all decrease. The adjusted $R^2$ and AIC of the two statistical models provide further evidence that even when the increased model complexity is taken into account, alcohol outlet visits still provide useful information to enhance the prediction of DV.

Table 6. A summary of the experiment results for testing the effectiveness of the derived alcohol outlet visits for enhancing DV prediction at the neighborhood level.

| | OLS | | RF | | DNN | | GWR | |
|---|---|---|---|---|---|---|---|---|
| | Baseline | Test* | Baseline | Test* | Baseline | Test* | Baseline | Test* |
| $R^2$ | 0.528 | **0.548** | 0.594 | **0.612** | 0.575 | **0.595** | 0.774 | **0.796** |
| RMSE | 9.653 | **9.452** | 8.978 | **8.848** | 9.250 | **9.018** | 6.681 | **6.352** |
| adjust-$R^2$ | 0.524 | **0.543** | - | - | - | - | 0.721 | **0.737** |
| AIC | 15861.358 | **15778.896** | - | - | - | - | 15056.949 | **14987.023** |

* Experiments that include alcohol outlet visits in the predictors for DV rate

The regression coefficients and feature importances derived from the OLS model and the RF model suggest that liquor store visit rate, in particular, is a critical factor for enhancing DV rate prediction at the neighborhood level. In the OLS model, *liquor store vr* is the largest positive predictor and the fourth largest predictor in terms of its absolute coefficient (following three racial and ethnic variables, which are *% White*, *% Hispanic*, and *% Asian*). In the RF model, *liquor store vr* is ranked as the 7th most important feature. The visit rates to drinking places and breweries also play fairly important roles in enhancing DV prediction.

Comparing across the four models in Table 6, we can see that GWR achieves the best prediction with an $R^2$ of 0.796. This is impressive given that the regular OLS model achieves an $R^2$ of only 0.548, and thus GWR provides about 45% improvement over the OLS model. OLS is a non-spatial model which examines the global effects of the independent variables on the dependent variable without considering local effects. The two advanced machine learning models, RF and DNN, improve the prediction over the OLS model, but they are also non-spatial models that do not take into account any spatial effect. The outstanding performance of GWR can be attributed to its ability to explicitly model spatial autocorrelation, a local effect we have observed in both neighborhood-level DV rate and the four types of alcohol outlet visit rates during the data exploration stage (see Figs 5 and 6).

We also plot out the predicted and observed DV rates of the four models in Fig. 9. As can be seen, the three non-spatial models (Figs 9(a), (b), and (c)) all underestimate neighborhood-level DV rate when the observed rate is higher than about 40, suggesting that a single global model does not provide good predictions for those neighborhoods with high DV rates. In contrast, the GWR model (Fig. 9(d)) greatly reduces those underestimates, and the prediction-observation data points in the scatter plot are not only located closer to the ideal diagonal line but also distributed more evenly along the two sides of the line. In addition, we plot out the standardized residual maps for OLS and GWR in Fig. 10, and calculate their global Moran's $I$ indices. As shown in the figure, the standardized residuals of the OLS model remain spatially autocorrelated, as indicated by its statistically significant and positive Moran's $I$ value. In comparison, the standardized residuals of the GWR model do not have a significant spatial autocorrelation.

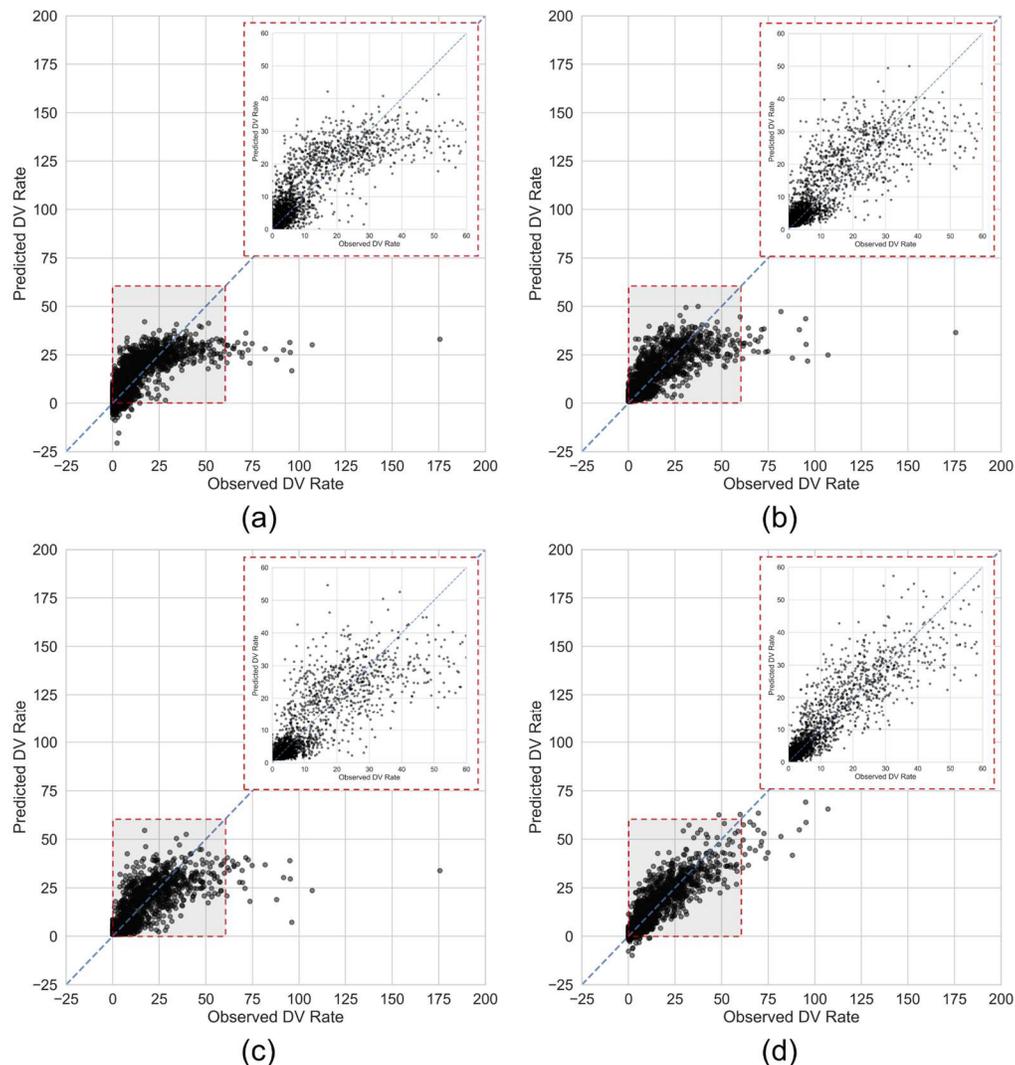

Figure 9. Predicted and observed DV rates based on the four models: (a) OLS; (b) RF; (c) DNN; (d) GWR. The top-right corners provide zoom-in scatter plots.

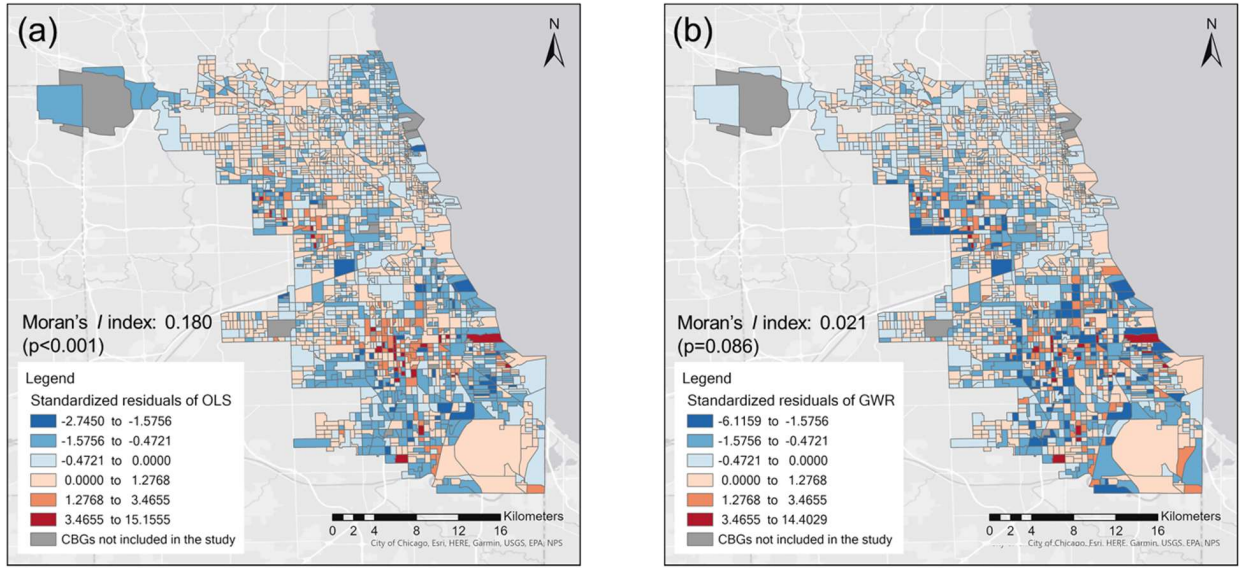

Figure 10. Standardized residual maps and their global Moran's *I* tests for (a) OLS; (b) GWR.

## 5. Discussion

### 5.1. The spatially varying role of alcohol outlet visits in DV prediction

The output of GWR includes local coefficients that enable us to examine the role of the four types of alcohol outlet visits across different geographic locations. We map out these local coefficients in Fig. 11. The cartographic technique Jenks Natural Breaks (Brewer & Pickle, 2002) is applied to each subfigure to classify and visualize the CBGs into six classes based on their local coefficients. Overall, these four maps show that alcohol outlet visits do not contribute to DV prediction in a homogeneous manner but have a spatially varying role across the city.

Figure 11(a) shows the effect of liquor store visits on DV rates. As can be seen, the local coefficients are mostly positive, suggesting that increases in liquor store visits are associated with increases in DV rates in most neighborhoods in Chicago. However, the extent of increase varies largely across the city, as shown in the different values of the local coefficients in the figure. Particularly, in the western and southern areas of Chicago, we observe clusters of high local coefficients ranging from 3.377 to 7.405, suggesting that in these neighborhoods visits of their residents to liquor stores are strongly associated with increased DV rates in these same neighborhoods. In contrast, in some other areas, such as the northern area of the city, we observe local coefficients that are close to 0, suggesting that visits to liquor stores have almost no effect on DV in these areas.

Figures 11(b) and (c) show the effects of drinking place visits and brewery visits on DV rates across the city. We observe mostly negative local coefficients, suggesting that increases in visits to drinking places or breweries are associated with decreases in DV rates in most neighborhoods. While surprising, this result is consistent with the results we obtained in the OLS model. In particular, we observe the strongest negative effects in the western and southern areas (i.e., the light color clusters in Figs 11(b) and (c)), which are roughly the same areas with the highest positive local coefficients for liquor store visits. These areas have low median household incomes based on their ACS data, and these different local coefficients seem to suggest that residents of these neighborhoods may visit liquor stores, rather than drinking places, to get alcohol with cheaper prices; meanwhile, visits to drinking places could be associated with relatively better socioeconomic statuses of some residents in these neighborhoods or certain good life events to celebrate, which could be linked to decreased DV rates. Such potential explanations, however, would need further investigation. In other areas of the city (i.e., the darker areas in Figs 11(b) and (c)), visits to drinking places or breweries have only slight or almost no effect on DV, as shown by the close to 0 local coefficients. Figure 11(d) shows the effect of winery visits on DV rates. However, we refrain from further interpreting the obtained coefficients given the small number of wineries (i.e., 16 wineries) in the study area.

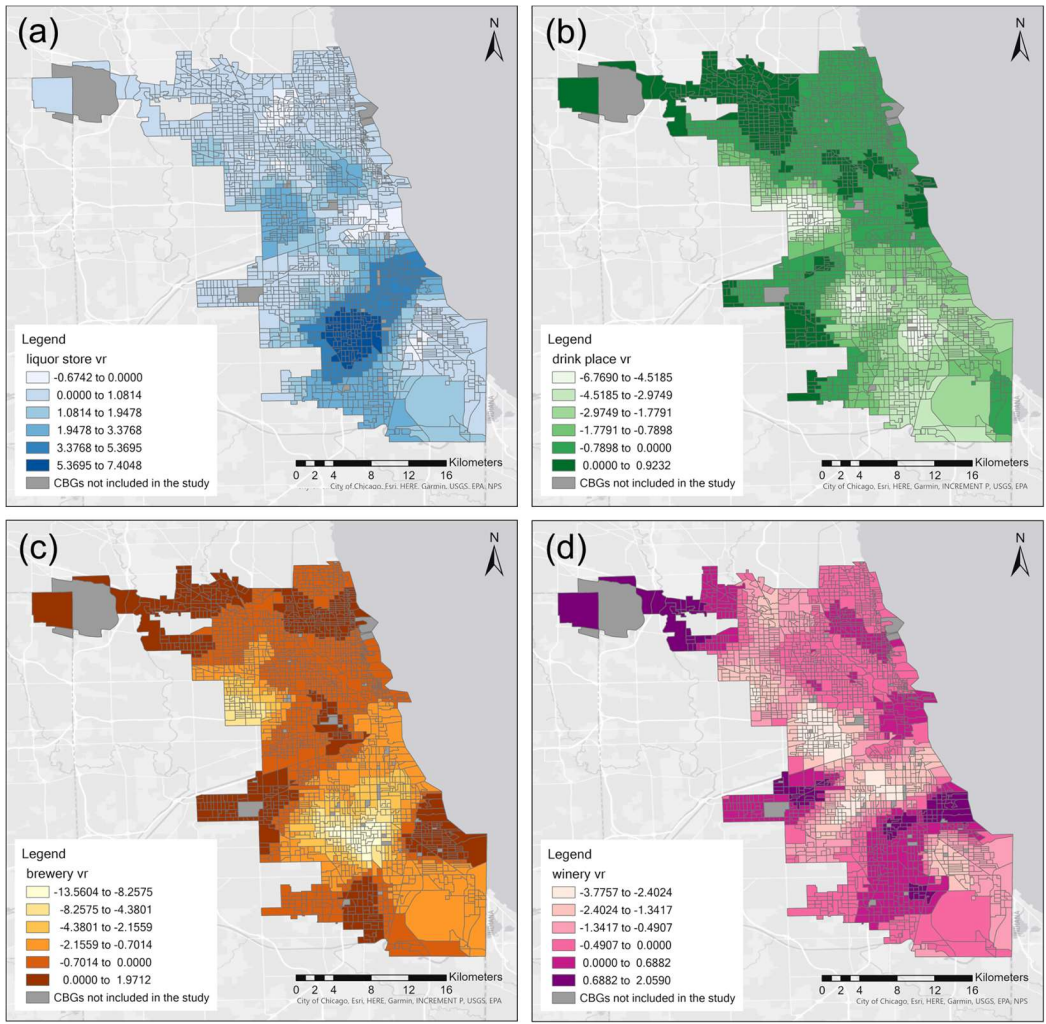

Figure 11. Local coefficients of the four types of alcohol outlet visits output by the GWR model: (a) liquor store vr; (b) drinking place vr; (c) brewery vr; (d) winery vr.

Our results provide further insights into the relationship between alcohol use and DV. Among the four types of alcohol outlet visits, liquor store visits have the strongest association with increased DV in the home neighborhoods of the liquor store visitors. The Routine Activity Theory of Crime (Cohen & Felson, 1979) suggests three necessary conditions for a crime to occur, which are: a motivated offender, a potential target, and a lack of anything to inhibit the offender's behavior. Liquor store visits suggest that the purchased alcohol will be consumed at home where all three conditions are met: the offender and the potential target are in close proximity to one another and the alcohol provides the disinhibitor (Giancola et al., 2011; Ito et al., 1996; Steele & Southwick, 1985; Testa et al., 2012). Although the way in which a DV incident occurs is likely to be more complex and may involve many other factors, such as gender difference (Cunradi et al., 2012), individual personalities, risky drinking behavior (Dey et al., 2014), and socioeconomic factors, the Routine Activity Theory provides a criminological perspective that helps to interpret our results.

## 5.2. Policy implications

Our study and the obtained results have important policy implications for governing bodies and public health officials working on DV intervention and prevention. We have shown that the inclusion of information about alcohol outlet visitation, which we derived from anonymized mobile phone location data, allows us to better identify the neighborhoods where DV is more likely to happen. DV intervention and prevention programs could then consider putting more effort into these neighborhoods. This is especially useful when the resources of DV intervention and prevention programs are limited, and focused effort on a smaller number of neighborhoods could have a higher positive impact than effort more evenly distributed throughout an entire city. From the perspective of informing alcohol licensing policies, the results from GWR demonstrate that the relation between alcohol outlet visits and DV exhibits spatial heterogeneity, suggesting that policymakers should more carefully examine the neighborhoods where residents' visits to alcohol outlets have stronger associations with increased DV incidents, and investigate why such stronger associations happen in these neighborhoods but not others. Policymakers could further identify the liquor stores that are usually visited by residents in these neighborhoods, and decide whether it is necessary to reduce alcohol availability in these places by, e.g., reducing their sales hours or by limiting the issuing of new alcohol licenses near these liquor stores (Stockwell & Gruenewald, 2004).

## 5.3. Limitations

This study is not without limitations. First, while the DV data are collected from an authoritative source (i.e., Chicago Police Department), they may contain data recording errors. As noted previously, some crimes not about DV were mistakenly labelled as "Domestic". Accordingly, it is also possible that some DV incidents are not correctly recorded. DV is also one of the most under-reported crimes (Beyer et al., 2015). Our current analysis therefore did not include these DV incidents that were either not recorded correctly or not reported. Second, to take into account spatial autocorrelation in the studied variables, we used the GWR model following the literature (Fotheringham et al., 2003; Weir, 2019; Oyana, 2020). However, other models, such as those based on Bayesian procedures, could be further explored to examine the effectiveness of the derived alcohol outlet visits for DV prediction. Third, while we have performed systematic experiments to test the effectiveness of the derived alcohol outlet visits for DV prediction, this study was based on the data from the city of Chicago. We can therefore extend this study to other cities or geographic regions where DV data are available, and to further test the effectiveness of alcohol outlet visits for DV prediction at the neighborhood level.

## 6. Conclusions

The paper investigates the role of alcohol outlet visits in enhancing DV prediction at the neighborhood level. We proposed a method for deriving neighborhood-level alcohol outlet visits by integrating anonymized mobile phone location data, POI data, and census data. Such a method can be applied to large geographic areas in a cost-effective manner. The derived alcohol outlet visits can represent how the residents of different neighborhoods visit various types of alcohol outlets throughout a city, and could be used as a proxy for neighborhood-level alcohol use. Because a visit to an alcohol outlet does not necessarily mean alcohol purchase, we examined the question "*Are alcohol outlet visits derived from anonymized mobile phone location data an effective predictor for DV at the neighborhood level?*" To answer this question, we conducted two sets of experiments using four different statistical and machine learning models to understand the role of the derived alcohol outlet visits in enhancing DV prediction. We used the DV data from the city of Chicago in 2019, and the experiment results showed that the derived alcohol outlet visits are indeed effective for enhancing DV prediction at the neighborhood level. However, the effect of alcohol outlet visits on DV is not homogeneous but varies spatially depending on the geographic areas within the city and the type of alcohol outlets. Our study was conducted at the CBG level, which can help DV intervention and prevention programs to identify the neighborhoods where DV is more likely to happen, and can help inform alcohol licensing decisions related to the liquor stores visited by residents of these neighborhoods. While this study has its limitations, it improves our understanding of the possibility to derive neighborhood-level alcohol outlet visits from anonymized mobile phone location data and the role of the derived alcohol outlet visits in enhancing DV prediction at the neighborhood level.